\documentclass[useAMS,usenatbib]{mn2e}
\input epsf.sty

\title{On the origin of bimodal duration distribution of Gamma Ray
	Bursts and the subjet model}
\author[A. Janiuk, B. Czerny, R. Moderski et al.]{A. Janiuk$^{1}$\thanks{E-mail:
agnes@camk.edu.pl}, B. Czerny$^{1}$, R. Moderski$^{1}$, D.B. Cline$^{2}$, C. Matthey$^{2}$, S. Otwinowski$^{2}$
\\
$^{1}$Nicolaus Copernicus Astronomical Center, Bartycka 18,
            00-716 Warsaw, Poland\\
$^{2}$University of California Los Angeles, Department of Physics and 
Astronomy, Box 951447, Los Angeles, CA 90095-1547, USA\\}

\begin{document}

\pagerange{\pageref{firstpage}--\pageref{lastpage}} \pubyear{2002}

\maketitle

\label{firstpage}

\begin{abstract}
The modified version of a bullet model for gamma ray bursts is studied. The central 
engine of the source produces multiple sub-jets that are contained within a cone. 
The emission
of photons in the source frame of a sub-jet either takes part in an 
infinitesimally thin 
shell, or during its expansion for a finite time.
The analysis of the observed profiles of GRBs taken by BATSE leads us to the conclusion
that the latter possibility is much more favored.
We also study the statistical distribution of GRBs, in the context of 
their bimodality of durations, taking into account the detector's capability of observing the signal above a certain flux limit. The model with shells emitting for a finite 
time is able to reproduce only one class of bursts, short or long, depending on
the adopted physical parameters. Therefore we suggest that the GRB bimodality is
intrinsically connected with two separate classes of sources. 
\end{abstract}


\section{Introduction}

The observed durations of Gamma-ray Bursts range from milliseconds to several hundreds
of seconds, forming two distinct peaks. Thus two GRB classes are proposed:
short ($T\le 2$ s) and long ($T\ge 2$ s) bursts (Kouvelietou et al. 1993).
(However, there have recently been claims of the third, intermediate
duration peak; see Horvath et al. 2004).
The collapsar scenario (Woosley 1993; Paczy\'nski 1998) 
is commonly favored for the origin of
a long GRB event, based on observations of afterglows and host galaxies
of these events (see the review by Zhang \& Meszaros 2004; Piran 2004). In particular,
in some long bursts  the 
afterglow observations clearly show the
association with supernovae (Stanek et al. 2003), 
thus confirming the collapsar scenario. 
Furthermore, the GRB positions inferred
from the afterglow observations are consistent with the GRBs being
associated with the star forming regions in their host galaxies.

Yet, the situation is still far from clear
for the short events, and various possibilities have been discussed.

An essentially different mechanism, such as a compact binary merger
(Eichler et al. 1989; Paczy\'nski 1991; Narayan, Paczy\'nski \& Piran 1992)
was proposed. The duration of the central
engine activity is in this case of the order of 2 seconds, appropriate
to account for the fueling time of a short GRB.
For many years, no optical counterpart was observed for a short burst but the
first detection of the X-ray afterglow by SWIFT satellite indicated
a location 9.8 arcseconds from the center of the elliptical (type E1) 
galaxy (Burrows et al. 2005) at a redshift of 0.2249 $+/-$0.0008 
(Prohaska et al. 2005). In such a galaxy strong starburst activity
is not expected, and no traces of the relative supernova were found
in the optical data (e.g. Kosugi et al. 2005) 
which again may favor a merger scenario. The results and implications for the burst
GRB 050509B were recently discussed in Gehrels et al. (2005).
For the recent observations of the GRB 050709, 
the source can be associated with the star forming galaxy at z=0.16 
(Fox et al. 2005; Villasenor et al. 2005), however with no
supernova associacion in the optical lightcurve (Hjorth et al. 2005).
Taking into account the distance, the luminosity of this
GRB afterglow is about 3 orders of magnitude lower than for a typical 
long event, making the coalescing compact binary the most promising progenitor
candidate (see also the discussion by Piro 2005).
Furthermore, recent observation of of GRB 050724 afterglow emission
supports strongly the merger origin of this event (Berger et al. 2005).

Also, the events of neutron star-neutron star (NS-NS) or neutron star-black hole 
(NS-BH) mergers should happen
quite often in the Universe (i.e. Bulik, Belczy\'nski \& Kalogera 2003;
Bulik, Gondek-Rosi\' nska i Belczy\' nski 2004).
These events should have given us a detectable sign of their presence,
not only in the form of gravitational waves, but also in the gamma-rays.

On the other hand, apparently the main argument in favor of
the merger scenario is that we cannot reject this possibility.
Therefore there have been attempts to unify one model
for all the GRBs, and assign all the bursts to the most plausible scenario, i.e.
collapsar.
One of the important arguments here 
 is the
 observation that a short GRB may be similar to the first $\sim
1$ s of a long one (Ghirlanda, Ghisellini \& Cellotti 2004).

It has been recently suggested (Yamazaki et al. 2004) that 
the bimodal duration distribution of GRBs can be explained in the
frame of inhomogeneous model of a GRB jet consisting of multiple
sub-jets, emitting for an infinitesimally short time.
Since the sub-jets, or sub-shells, are not distributed uniformly
within the main jet, but instead may have e.g. Gaussian (or power-law)
distribution depending on the angular distance from the jet axis,
the observer does not always detect the same number of sub-jets
on his line of sight. In other words, the number of sub-jets, $n_{\rm s}$,
seen by the observer depends on the viewing angle. 

The multiple sub-jet model proposed by
Yamazaki et al. (2004) assumes that the duration of the observed
burst does not essentially depend on the duration of activity of the
central engine. Or, at least, the central engine of both short and
long bursts may be of the same nature, i.e. a collapsar, operating
for, say, 30 seconds. It means that duration of activity of the central
engine is only slightly reflected in the duration of the observed GRB
event. Instead, the event duration depends mainly on the viewing angle. This is
because the observer may either detect a large number of sub-jets
($n_{\rm s}>1$), and in consequence the burst duration is long, or the
number of sub-jets detected on the line of sight is $n_{\rm s}=1$ and we
observe a short burst.

In this article we build a similar
 model of a non-uniform jet, that consists of a
number of randomly distributed sub-jets. We study both the case of sub-jets
emitting the gamma rays 
at an instantaneous time $t=t_{0}$, as invoked by
Yamazaki et al (2004), and the case of
a more realistic assumption that 
the emission of gamma rays is not instantaneous in time, but lasts
for a certain period during the expansion of the shell from $R_{\rm min}$ to
 $R_{\rm max}$. 
We calculate the profiles of individual sub-bursts as well as the
total lightcurves of the GRBs that are the sum of multiple
sub-bursts.
Next, we fit the profiles of the individual pulses using the
phenomenological prescription of Ryde \& Svensson (2002)
and check if these profiles correspond to these
seen on the observational data from BATSE. 
The data were analyzed both by  Ryde \& Svensson (2002) 
in case of long duration events
and by ourselves in case of short events.

We also simulate the statistical  distribution
of durations of the GRBs.
Firstly, we hold the assumption, that only the sub-jets present
on the line of sight of the observer account for the GRB event.
Secondly, we release this assumption, and include also the emission
from the sub-jets detected off-axis.
The statistic is calculated on the basis of the 
condition used for classifying the events as GRB, i.e. the flux limit 
(the detector's capability).   

In Section \ref{sec:model} we give the details of the model and assumptions.
In Section \ref{sec:modprofiles} we present the resulting profiles of the bursts and 
sub-bursts and in Section \ref{sec:obsprofiles} we compare 
these profiles with observations.
In Section \ref{sec:statistics} we show the statistical distributions 
of the burst durations that
result from the multiple sub-jet model.
Finally, in Section \ref{sec:diss} we discuss our results and 
give conclusions.

\section{Model}
\label{sec:model}

\subsection{Geometry}
\label{sec:geom}

We consider a collimated jet in the form of a cone with the opening angle
$\Delta \theta_{\rm tot}$, which is present for the whole period of the
duration of central engine activity, $T_{\rm dur}$. 
The main jet consists then of a number
$N_{\rm sub}$ of sub-jets, which have smaller opening angles  
$\Delta \theta_{\rm sub}$ and are launched at their characteristic
departure times, extending until the time $T_{\rm dur}$:  $0<t_{\rm dep}<
T_{\rm dur}$. The position of each sub-jet is determined by two angles,
($\theta_{\rm i}$, $\varphi_{\rm i}$), such that the angular distribution
of the direction $\theta_{\rm i}$, i.e. the distance from the $z$ axis of
the main jet, is given by the Gaussian function:
\begin{equation}
P(\theta_{\rm i}) d\theta_{\rm i} = e^{-{1\over 2}{\theta_{\rm i} \over
\theta_{\rm c}}^2} d\theta_{\rm i}, ~~~~ for~~~ \theta_{\rm i}<\Delta\theta_{\rm tot}-\Delta\theta_{\rm sub}
\end{equation}
where $\theta_{\rm c}<\Delta \theta_{\rm tot}$ is a parameter of this distribution.
Independently, the angle $\varphi_{\rm i}$, measured in the $x-y$ plane in
the basis of the jet, is uniformly distributed between 0 and 2$\pi$.

We assume that all the subjets have the same Lorentz factors $\gamma$
and up to the radius $r_{0}$, for a period of time $t_{0}$, expand in
a photon-quiet phase. At the radius $r_{0}$ the photons are emitted 
instantaneously, i.e. the duration of the photon-active phase in the
comoving frame is infinitesimally small. The sub-jet $i$ 
is expanding for the time:
\begin{equation}
t_{0}^{i} = t_{\rm dep}^{i} + {r_{0}\over c \beta}
\end{equation}
where $\beta = \sqrt{1-1/\gamma^{2}}$. Then, all the photons are
emitted in an infinitesimally short time, at:
\begin{equation}
t_{\rm em}^{i} = {r_{0}\over c \beta}
\end{equation}
after the launch of a sub-jet.

The observer's position is determined by his viewing angle ($\theta_{\rm obs}$,
$\varphi_{\rm obs}$), as measured in the reference frame of the main jet. 
Again, we randomly choose these angles, without favoring any
particular direction, and the angles are $0<\theta_{\rm obs}<\Delta
\theta_{\rm tot}$ and $0 < \varphi_{\rm obs} < 2\pi$. The observed
temporal structure of a flash of radiation produced by a sub-jet
is determined by the time measured in the observer's frame,
i.e. the arrival time of a photon:
\begin{equation}
T_{\rm obs}^{i} = (1+z)(t_{0}^{i} + t_{\rm em}^{i}(1-\beta\cos\lambda))
\end{equation} 
where we take into account that relativistic aberration of photons
emitted at an angle $\lambda$ to the observer occurs
only after the expansion time $t_{0}^{i}$. This is because
 the assumed radius of the emitting photosphere is large and cannot 
be neglected. Here the angle $\lambda$ is measured between the
direction of motion of the emitting plasma with respect to the
observer. 

\begin{figure}
\epsfxsize = 250pt
\epsfbox{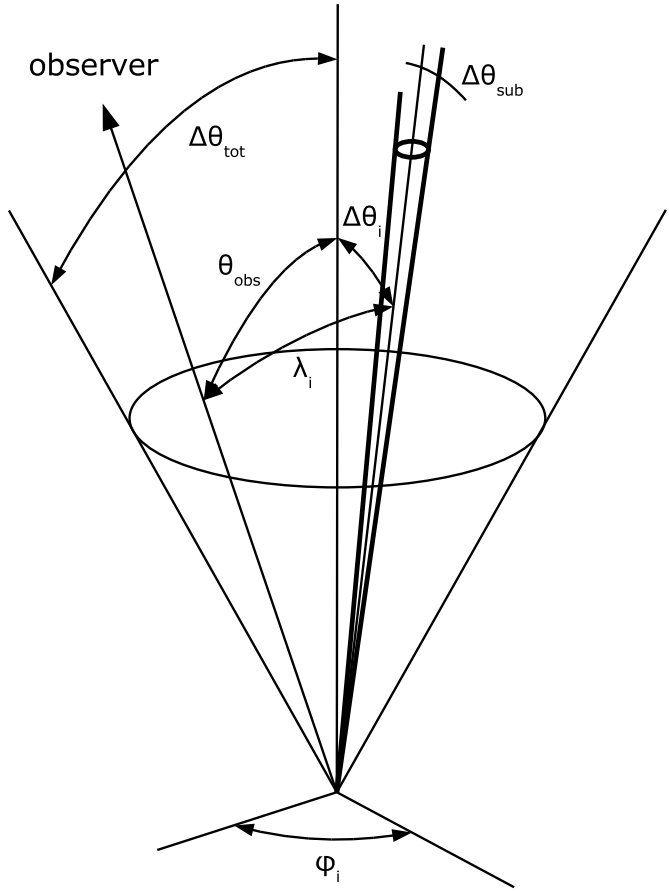}
\caption{Geometry of the problem.  The jet has a half-opening angle
$\Delta\theta_{\rm tot}$.  The observer is located at ($\varphi_{\rm obs} = 0$,
$\theta_{\rm obs}$) in the jet reference frame.  Each subjet has a
half-opening angle $\Delta\theta_{\rm i}$ and is located at
$(\varphi_{\rm i},\theta_{\rm i})$.  The angle between the observer and the sub-jet
axis, $\lambda_{\rm i}$, form a spherical triangle together with the angles
$\theta_{\rm obs}$ and $\theta_{\rm i}$.
\label{fig:geom}}
\end{figure}

Because of the axial symmetry of the problem we choose an observer to be
located at $\varphi_{\rm obs} = 0$ in the reference frame of the jet.
In order to determine this angle $\lambda$ 
and its dependence on the observer's
position, location of the subjet, and direction of the photon,
let us consider a spherical triangle located on the surface of the sphere
of the radius $r_{0}$. The vertexes of this triangle are determined by 
the directions of the jet axis, observer's axis, and sub-jet axis.
Therefore the sides of the triangle are $\theta_{\rm obs}$, $\theta_{\rm i}$, 
and $\lambda_{\rm i}$, while the angle opposite to $\lambda_{\rm i}$ is
$\varphi_{\rm i}$ (see Figure~\ref{fig:geom}).

This implies the dependence:

\begin{equation}
\cos \lambda_{\rm i} = \cos\theta_{\rm i}\cos\theta_{\rm obs} +
\sin\theta_{\rm i}\sin\theta_{\rm obs}\cos(\varphi_{\rm obs}-\varphi_{\rm i})
\label{eq:spheric}
\end{equation}

Now, the maximal and minimal angle $\lambda$ between the
direction of motion of the emitting plasma with respect to the
observer,  is given by:
\begin{eqnarray}
\label{eq:lambdamin}
\lambda_{\rm max} = \lambda_{\rm i}+\Delta\theta_{\rm sub} \\
\nonumber
\lambda_{\rm min} = min(0; \lambda_{\rm i}-\Delta\theta_{\rm sub})
\end{eqnarray}

Having calculated the angle $\lambda_{\rm i}$, we can determine the time $T_{\rm obs}$ at which the photon hits the detector. The duration of a single sub-burst is therefore given by the difference between the ending and starting times in the observer's frame:

\begin{eqnarray}
T^{i}_{\rm start} = T^{i}_{0} + (1+z){r_{0} \over c\beta}(1 - \beta\cos\lambda_{\rm min}) \\
\nonumber
T^{i}_{\rm end} = T^{i}_{0} + (1+z){r_{0} \over c\beta}(1 - \beta\cos\lambda_{\rm max})
\end{eqnarray}
where $T^{i}_{0}=(1+z)t^{i}_{0}$.

In the above considerations, for simplicity we neglect the dependence
of the angle $\lambda$ between photon's direction and observer's position on 
the angle $\varphi_{\rm obs}$. This is because we have random observers
uniformly distributed between 0 and $2\pi$, so for a sample with large 
enough statistics this will not influence the results.
Also, please note that we did not give an exact formula for the angle $\lambda$
but rather its upper and lower limits resulting from the opening angle of
a sub-jet. The derivation of an exact expression would require
a transformation between the two reference frames: jet's (in which 
the observer's position is given) and sub-jet's (in which the photons 
from a single sub-jet are produced).

Therefore the simplified
 condition for the observer's position  to fall 
inside the sub-jet is:
\begin{equation}
\lambda_{\rm i} \le \Delta\theta_{\rm sub}.
\label{eq:inside}
\end{equation}
and also neglects the dependence on the angle $\varphi$.

\subsection{Duration of a burst}

If the line of sight of the observer coincides with only one sub-jet, then the 
duration of this sub-jet is equivalent to the duration of the GRB.
Alternatively, the observer may 'catch'  a number of sub-jets. 
In this case the duration of the GRB is the sum of durations of all the 
sub-bursts that coincide with the  line of sight:
\begin{equation}
\Delta T = Max(T^{i}_{\rm end}) - Min(T^{i}_{\rm start})
\end{equation}

The absolute maximal and minimal duration of a GRB event are estimated as follows.
The duration of a single sub-burst for $\lambda_{\rm i}>\Delta\theta_{\rm sub}$ is:
\begin{eqnarray}
c \Delta t = r_{0}(\cos(\lambda_{\rm i}-\Delta\theta_{\rm sub}) - \cos(\lambda_{\rm i}+\Delta\theta_{\rm sub})) \\
= 2 r_{0} \sin\lambda_{\rm i}\sin\Delta\theta_{\rm sub} \nonumber
\end{eqnarray}
while for  $\lambda_{\rm i}<\Delta\theta_{\rm sub}$ it is:
\begin{equation}
c \Delta t = r_{0}(1-\cos(\lambda_{\rm i}+\Delta\theta_{\rm sub}))
\end{equation}
(cf. Eq. \ref{eq:spheric}).
Therefore the shortest possible sub-bursts are these seen on-axis ($\lambda_{\rm i}=0$), and:
\begin{equation}
\Delta t_{\rm min} \approx {r_{0} \over c} {1 \over 2}\ (\Delta\theta_{\rm sub})^{2}
\label{eq:tmin}
\end{equation}
For our standard parameters, $r_{0} = 10^{14}$ cm and 
$\Delta\theta_{\rm sub}=0.02$, equation (\ref{eq:tmin}) gives 
$\Delta t_{\rm min} = 0.66$ s.

The longest  theoretically 
possible burst would be seen at the angle $\lambda_{\rm i}=\pi/2$,
and when the sub-bursts caught on the line of sight were so numerous that they 
would cover the whole jet opening angle: instead of $\Delta\theta_{\rm sub}$ we
take $\Delta\theta_{\rm tot}$. In case of our parameters ($\Delta\theta_{\rm tot}=0.2$,
$T_{\rm dep} = 30$s)
this is about 1350 s.
 In practice, as discussed in the next section, the light aberration will
effectively limit the observed surface to within a small angle around
the line of sight. In such a case the duration of the longest burst will
strongly depend on the central engine lifetime $T_{\rm dep}$ and on the
sensitivity of the instrument used to observe the burst."

\subsection{Sub-burst profiles}
\label{sec:profiles}

The duration of the GRB event depends not only on the number of sub-jets
that are observed, but also on the energy flux that is measured by the 
detector. This is of particular importance if we allow also for the sub-bursts 
that are observed off-axis, to contribute to the observed spectrum. Because
the energy 
flux observed from the relativistic jet sharply decreases for the viewing angles
that are outside the jet cone, only a part, if at all, of such a sub-burst could
be classified as a gamma ray burst.

The energy flux emitted from an expanding shell, 
 as seen by the distant observer located on the jet axis, 
was studied in Fenimore at el. (1996). They considered both the case
of an infinitesimally thin and  extended shell. 

The photons emitted from a relativistically expanding shell, 
that arrive to the observer at 
the same time, originate from a prolate ellipsoid (Rees 1966). In order to calculate 
the flux in the detector rest frame, we have to integrate the emitted spectrum
over the surface 
resulting from the cross-section of the
ellipsoid with the sphere of the center in the detector and radius equal to the 
distance to the source:

\begin{equation}
dF_{\nu}(T)  \propto \oint f(\nu') \delta^{3} dA
\label{eq:flux}
\end{equation}
where $\delta = [\gamma (1-\beta \cos \theta)]^{-1}$ is the Doppler factor, $f(\nu')$
is the emissivity
 in the comoving frame, and $\nu'=(1+z)\nu\gamma(1-\beta \cos \theta)$.
The surface differential is determined by:
\begin{equation}
dA = {1 \over D^{2}} \Delta \Phi \sin\theta d\theta 
\end{equation} 
where
\begin{eqnarray}
\Delta \Phi = \pi ~~~~ if ~~ \theta < (\Delta\theta_{\rm sub}-\theta'_{\rm obs}) ~~or~~  \theta_{\rm obs}=0.  \nonumber  \\
\Delta \Phi = 2 \arccos({\cos\Delta\theta_{\rm sub}-\cos\theta \cos\theta'_{\rm obs}\over 
\sin\theta \sin\theta_{\rm obs}}) ~~~ otherwise
\end{eqnarray}
 
The angle $\theta'_{\rm obs} \equiv \lambda_{\rm i}$ is between the direction to the observer and the sub-jet axis. The angle $\theta$ 
depends on time in the observer's frame, and here is how the observed pulse varies 
with time:
\begin{equation}
\theta(T) = \arccos({1 \over \beta}-{1\over 1+z} {c\over r}(T-T_{0}))
\label{eq:theta}
\end{equation}
We calculate the pulse profile for $T_{\rm start}<T<T_{\rm end}$, 
and in principle we should integrate
from $\theta_{\rm min}$ to $\theta_{\rm max}$ (where $\theta_{\rm min} = 0$
if $\theta'_{\rm obs}<\Delta\theta_{\rm sub}$ or $\theta_{\rm min} = \theta'_{\rm obs}-\Delta\theta_{\rm sub}$ otherwise, and $\theta_{\rm max} = \theta'_{\rm obs}+\Delta\theta_{\rm sub}$).
However, in case of an infinitesimally thin shell, we must keep $r=r_{0}$
and only one value of $\theta$ will satisfy the relation \ref{eq:theta}.
In other words, the integral in equation \ref{eq:flux}
becomes one-dimensional integral over the arc of the length $\Delta\phi$.

\subsection{Shell emission for a finite time in the comoving frame}

The model described above considered the emission that is infinitesimally short in the
comoving frame, and the duration of the observed pulse is a result 
of the purely geometrical transformation to the observer's frame.
In reality, the emission from the expanding shell may not be instantaneous, 
but rather
lasts for a certain time. In our model the duration of the emission is determined
 by the minimum and maximum radii of the expanding shell, $r_{\rm min}$ and $r_{\rm max}$.
The starting and ending time of an individual sub-burst in the observer's frame 
is given by:
\begin{eqnarray} 
  T^{i}_{\rm start} = (1+z)(t^{i}_{\rm dep}+{r_{\rm min}\over c \beta}(1-\beta \cos\lambda_{\rm min}))\\
  T^{i}_{\rm end} = (1+z)(t^{i}_{\rm dep}+{r_{\rm max}\over c \beta}(1-\beta \cos\lambda_{\rm max}))\nonumber
\end{eqnarray}
where $\lambda_{\rm min}$ and $\lambda_{\rm max}$ are given by the equations \ref{eq:lambdamin}.
In order to obtain the time profile of each sub-burst, we have to calculate
the integral in Eq. \ref{eq:flux}, over $\theta$ and $\Delta \phi$, where
the limits for $\theta(T)$ will satisfy the condition $r_{\rm min}<r<r_{\rm max}$.
In this way we sum up in the observer's frame all the photons
that are emitted in different moments in the expanding shell between $r_{\rm min}$ and $r_{\rm max}$ but arrive to the detector at the same time T.

\subsection{Comparison with previous work}

The idea of the GRB jets consisting of multiple sub-jets was proposed by Heinz \& 
Begelman 
(1999), and recently modeled in detail by Yamazaki et al. (2004) and 
Toma et al. (2005). In the latter work the authors successfully reproduced the bimodal 
distribution of the GRB durations.
Here we basically follow the same scheme of the multiple sub-jet model, however 
there are a few differences. 

(i) First, we calculate the angles between 
the emitting electrons and observer's direction, explicitly taking into account 
the sub-jet azimuthal angle, $\varphi_{\rm i}$ 
(see equation \ref{eq:spheric} and Fig. \ref{fig:geom}). 

(ii) Secondly, we distinguish between the cases with and without the off-axis emission 
included in the total GRB pulse. This is particularly important, when determining the 
duration of a short burst and statistical fraction of short bursts in the simulation.

(iii) Therefore, thirdly, we adopt a different way of selecting the pulses 
that are responsible for gamma ray signal. Instead of the spectral hardness,
we adopt a criterion based on the  sufficient signal with respect to the background 
flux. The minimum flux is inversely proportional to the square root of $T_{90}$, as 
implied by the BATSE characteristics (see Section \ref{sec:offaxis}).

(iv) Finally, we  include in the model the possibility of a 
non-instantaneous emission, 
in the expanding shell that emits radiation from $R_{\rm min}$ to $R_{\rm max}$.

\section{Results}
\label{sec:results}

\subsection{Pulse profiles}
\label{sec:modprofiles}
We present here the shapes of the profiles for individual sub-bursts
as well as the total GRB events that consist of multiple sub-jets.
First, we show the results for the case of an infinitesimally thin emitting shell.
In this case the emission in the source frame is instantaneous, and the
duration of the event is determined only by the geometry.

\subsubsection{Profiles for the infinitesimally thin shell emission}

\begin{figure}
\epsfxsize = 250pt
\epsfbox{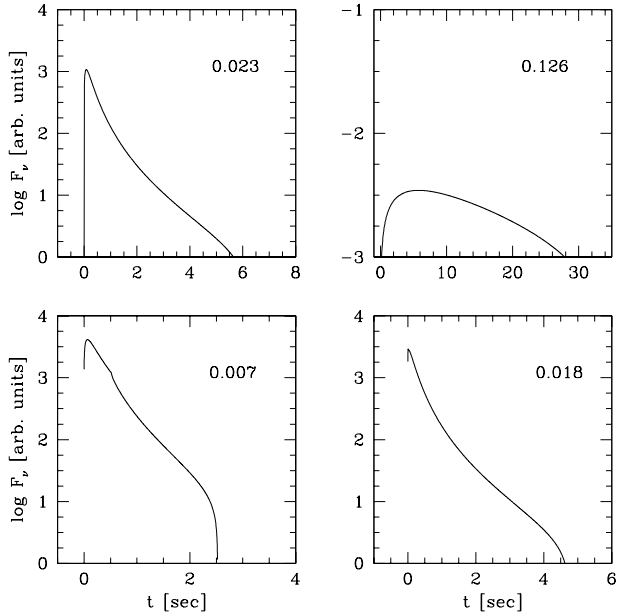}
\caption{The exemplary pulse profiles emitted from the individual sub-jets
as seen by the observers located at various lines of sight.
The angle $\theta'_{\rm obs}=0.007, 0.018, 0.023$ and 0.126 is indicated
in  each panel. The sub-jet opening angle is always
$\Delta \theta_{\rm sub}=0.02$ and the Lorentz factor is $\gamma = 100$.
The flux is given with arbitrary normalization and essentially does not
depend on frequency (flat spectrum).
\label{fig:Spectab}}
\end{figure}

In Figure \ref{fig:Spectab} we show the exemplary profiles 
of the pulses from the individual sub-jets, as seen by the 
observer.
The sub-jet opening angle is $\Delta \theta_{\rm sub}=0.02$
and therefore the pulses that are seen 
inside this opening angle are much more energetic.
The pulses seen off-axis are weaker, due to the strong dependence
of the flux on the Doppler factor and in turn on the viewing angle.
This means that the maximum of the observed flux is almost an order of
 magnitudes lower,
as soon as the observer is located outside $\Delta \theta_{\rm sub}$
(see e.g. cases of $\theta'_{\rm obs}=0.018$ and $\theta'_{\rm obs}=0.023$).
On the other hand, for larger viewing angles the pulse duration, as well
as its starting time, 
substantially increase. 
 For $\theta_{\rm obs} = 0.007$ a break in the
light curve is caused by the conical geometry of the jet as discussed by
Rhoads (1997).

\begin{figure}
\epsfxsize = 250pt
\epsfbox{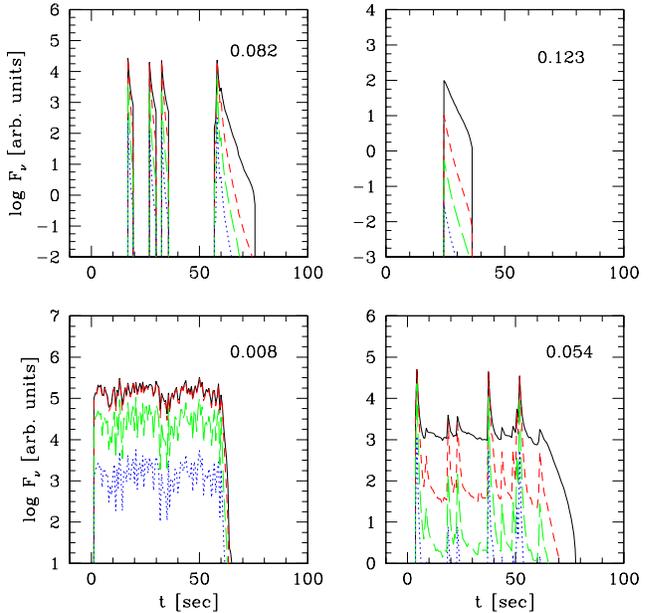}
\caption{Profiles of the GRBs for  various 
observing angles with respect to the main jet axis:
$\theta_{\rm obs}$=0.008, 0.054, 0.082 and 0.123. 
The lightcurves are plotted in 4 energies: 1.41 (solid line), 112 (short dashed line),
 482 (long-dashed line) and 2070 keV (dotted line).
The bursts contain the total emission (on and off-axis) from the sub-jets,
emitting at $r_{0}=10^{14}$ cm, that satisfy the condition
Eq. \ref{eq:fluxlimit} (see text).
\label{fig:r0sum}}
\end{figure}

In Figure \ref{fig:r0sum} we show the total profiles of GRB, which
are the result of multiple pulses seen by the observer.
The plots are labeled here with the observer angle with respect to the
 axis of the jet. While looking at small $\theta_{\rm obs}$, the observer
will detect a large number of sub-pulses that overlap, and give rise to a
long, variable GRB.
In contrast, the observer looking at larger angle $\theta_{\rm obs}$,
can detect only a few sub-pulses. These form either a set of separate
narrow spikes, or are connected with each other by the emission 
from these sub-bursts that were seen only off-axis
(see Section \ref{sec:offaxis} and condition Eq.
\ref{eq:fluxlimit} for the detection of the off-axis emission).

The spectrum in the comoving frame is defined using the 
broken power law function (see also the formula in Band et al. 1993):
\begin{eqnarray}
\label{eq:brokenpow}
f_{\nu'} = (\nu'/\nu_{0}')^{1+\alpha} ~~~~ for ~~~~~ \nu'< \nu_{0}' \nonumber \\
 ~~~~~~(\nu'/\nu_{0}')^{1+\beta} ~~~~ for ~~~~~~ \nu'> \nu_{0}'
\end{eqnarray} 
In the comoving frame the
  break frequency is assumed $\nu_{0}'= 2keV$, and the spectral indices are
$\alpha = -1$ and $\beta=-3$ (the averaged values of the observed spectral indices were
given e.g. by Pendleton et al., 1994; Preece et al. 1998, 2000).
The lightcurves in Fig. \ref{fig:r0sum} 
are plotted in 4 energies: 1.41,112, 482 and 2070 keV.

\subsubsection{Emission from the thick expanding shell}

Now, we show the results for the case of a thick expanding shell.
In this case the duration of the individual pulse is determined not only by the 
geometry but also by the duration of the emission in the source frame.

\begin{figure}
\epsfxsize = 250pt
\epsfbox{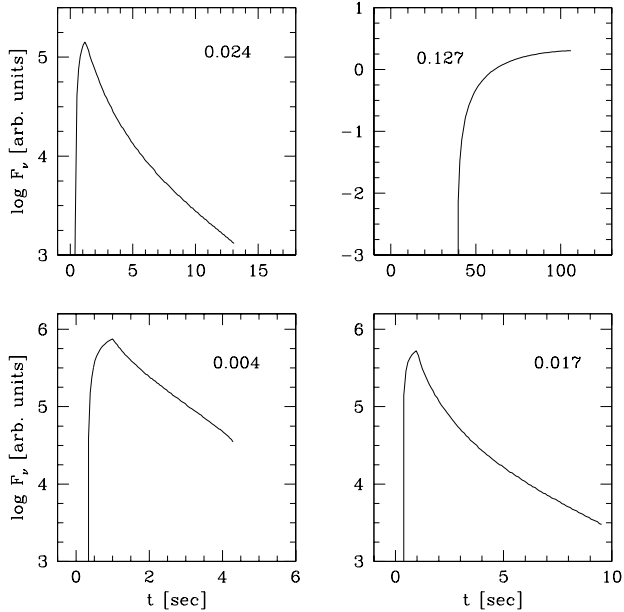}
\caption{Profiles of the individual sub-bursts for  various 
observing angles with respect to the sub-jet axis:
$\theta'_{\rm obs}$=0.004, 0.017, 0.024 and 0.127. Thick expanding shell approach. 
\label{fig:flaretab}}
\end{figure}

In Figure \ref{fig:flaretab} we plot several examples of the
 individual sub-bursts, for various 
observing angles with respect to the sub-jet axis:
$\theta'_{\rm obs}=0.004, 0.017, 0.024$  and 0.127. 

\begin{figure}
\epsfxsize = 250pt
\epsfbox{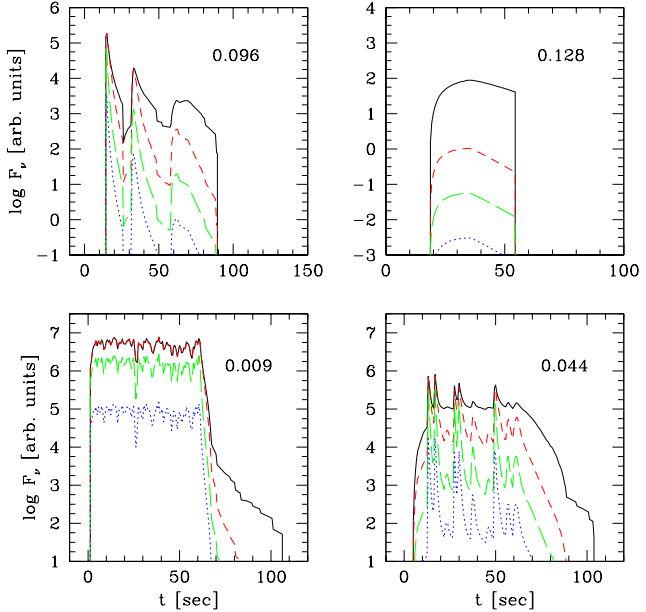}
\caption{The same as in Fig. \ref{fig:r0sum}; the 
observing angles are:
$\theta_{\rm obs}$=0.009, 0.044, 0.096 and 0.128. 
The bursts contain the total emission (on and off-axis) from the sub-jets,
from $R_{\rm min}=10^{14}$ cm to $R_{\rm max}=2\times 10^{14}$ cm.
\label{fig:flaresum}}
\end{figure}

In Figure \ref{fig:flaresum} we plot the total GRBs, consisting of multiple sub-pulses,
for several exemplary values of the observers angle (with respect to the main jet
axis): $\theta_{\rm obs}=0.009, 0.044, 0.096$ and 0.128. The Figure shows the lightcurves in 4 energies: 1.41, 112, 482 and 2070 keV. 
The jet opening angle is $\Delta\theta_{\rm tot}=0.2$.
For the bursts that are observed close to the edge of the cone (large
$\theta_{\rm obs}$), the probability of detecting a sub-jet on-axis is very small. These 
bursts consist mostly of the off-axis emission from the sub-jets, typically either 
with one-two strong, on-axis pulses (case of $\theta_{\rm obs}=0.096$) or
with none on-axis pulse (case of $\theta_{\rm obs}=0.128$).
These bursts have the longest total durations-and last for several hundreds of seconds.
The bursts that are observed closer to the main jet axis, are stronger and
 more variable,
and their durations typically range up to $\sim$ 100 seconds. 

\subsection{Fitting the pulse profiles}
\label{sec:obsprofiles}

The observed profiles of GRBs can be fitted to the exponential function in the rise phase:
\begin{equation}
N(t) \propto \exp(m t)
\end{equation}
and to the power law shape in the decay phase (Ryde \& Svensson 2002):
\begin{equation}
N(t) \propto (1+t/\tau)^{n}
\label{eq:funsvensson}
\end{equation}
where $m, n$ and $\tau$ are the free parameters.
As shown by Ryde \& Svensson (2002) for a sample of BATSE data, 
the distribution of the index $n$, characteristic 
for the pulse decay phase, peaks twice: around $n=1$ and $n=3$. However, one should 
keep in mind that this observational result has been obtained using a limited 
statistics of $\sim 20$ long GRBs detected by BATSE.
Here we fit the pulse profiles that are obtained from the sub-jet model.

\begin{figure}
\epsfxsize = 250pt
\epsfbox{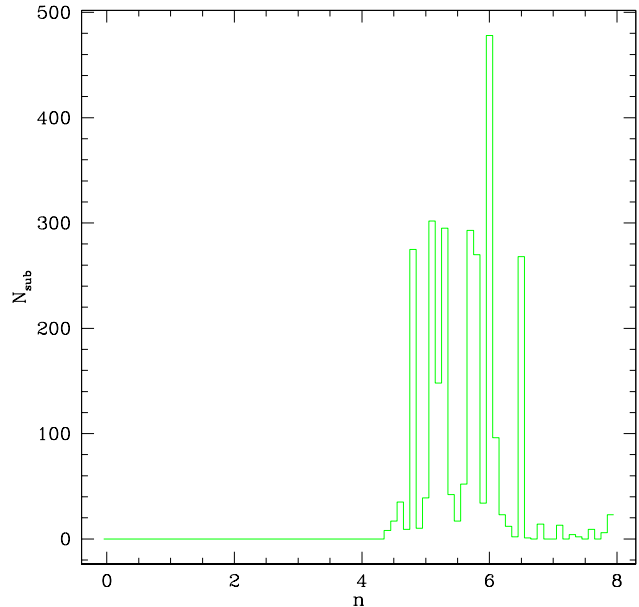}
\caption{The distribution of the index $n$, fitted to the sub-pulses profiles
due to the instantaneous emission at $r=r_{0}$.
\label{fig:wyk2}}
\end{figure}

First, we fit the pulses resulting from the assumed delta-function emission 
($t=t_{0}$,  $r=r_{0}$) in the comoving frame.
In the Figure \ref{fig:wyk2} we show the distribution of indices for 
the sub-pulses that fitted to this pulse profile, 
and for which the maximum gamma-ray flux was substantially large (this corresponds
roughly to the observing angle $\theta'_{\rm obs}>0.03$).
The indices in our distribution are in the range $n=4-6$, while neither
$n=1$ nor $n=3$ showed up, contrary to the results obtained by Ryde \& Svensson (2002).

\begin{figure}
\epsfxsize = 250pt
\epsfbox{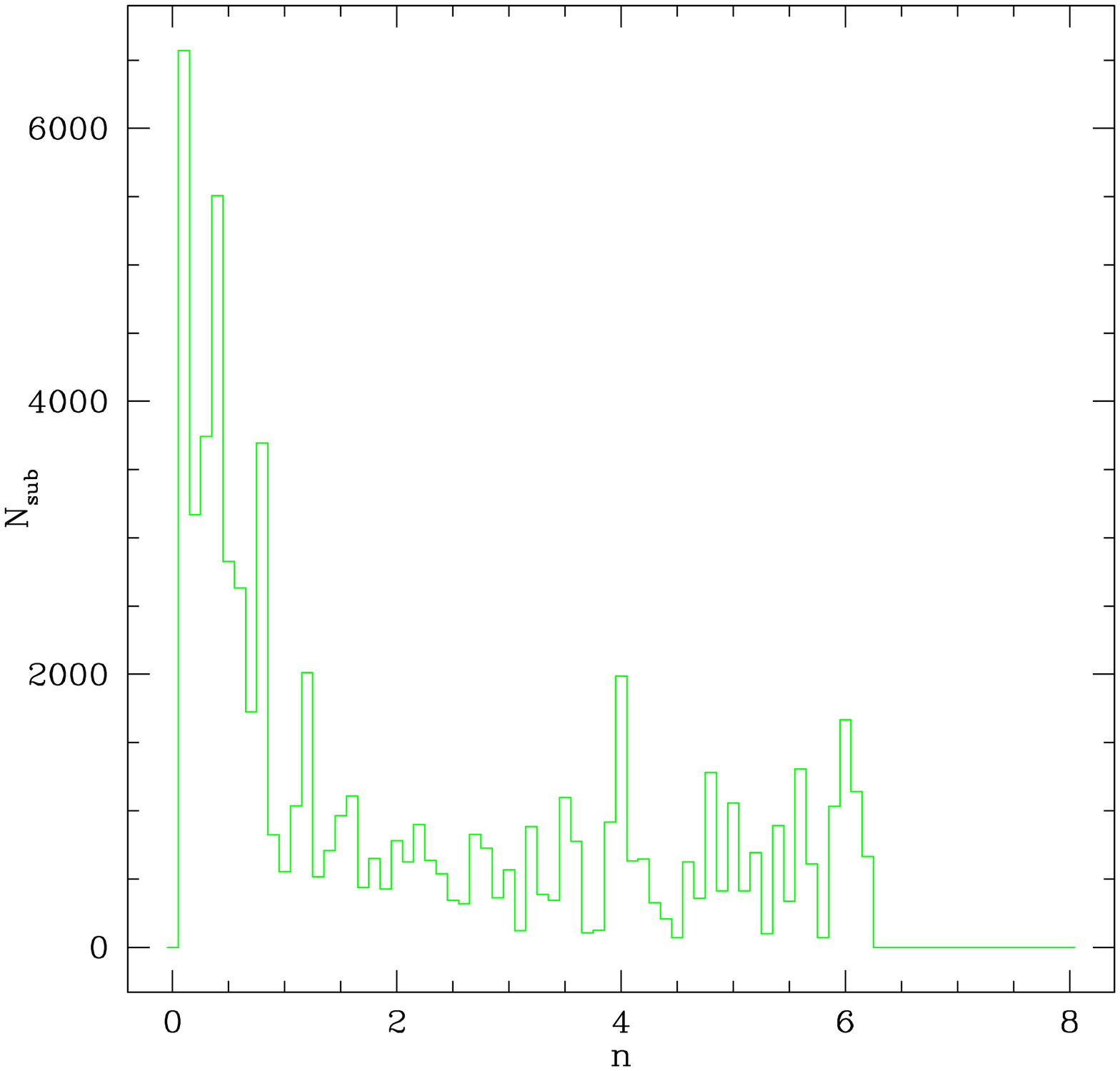}
\caption{The distribution of the index $n$, fitted to the sub-pulses profiles
due to the emission from the expanding shell from $R_{\rm min}=10^{14}$ cm to 
$R_{\rm max}=2\times 10^{14}$ cm.
\label{fig:wyk2fl}}
\end{figure}

Next, we fit the pulses resulting from the assumed  emission
in the expanding shell from $R_{\rm min}$ to $R_{\rm max}$.
In the Figure \ref{fig:wyk2fl} we show the distribution of indices for all 
the sub-pulses that fitted to this pulse profile (again 
apart from the sub-jets seen at  angles $\theta'_{\rm obs}>0.03$).
The distribution is now much broader, with certain enhancement at low values of 
$n$, but with no traces of $n=3$ secondary peak, as found by 
Ryde \& Svensson (2002).

\subsubsection{New sample from BATSE}

Here we present the analysis of the new sample of BATSE data, 
in order to
make further observational tests for the model.
Now, contrary to Ryde \& Svensson (2002), we study only the short duration 
bursts.
The sample of bursts is listed in Table \ref{tab:batse} and
 contains 83 GRBs, of durations from 0.067 s to 1.954 s.
The data from the energy band 50 -- 300 keV were used and the bursts were 
selected so that the signal to noise ratio in this band was larger than 10.

\begin{table*}
\begin{tabular}{|l|c|c|c|c|r|c|l|c|c|c|c|r|}
\hline
Trigger & $T_{90}$ & S/N & $\chi^{2}$ & $\tau$ & $n$ &  & Trigger & $T_{90}$ & S/N & $\chi^{2}$ & $\tau$ & $n$ \\ 
\hline

    207 &  0.085 & 16.77 &  1.83 &  0.016 &  2.43 &  & 3502 &  0.416 & 19.62 &  1.01 &  0.011 &  0.93 \\
    480 &  0.067 & 31.04 &  1.44 &  0.046 &  4.50 &  & 3644 &  0.776 & 38.84 &  1.22 &  0.276 &  3.18 \\
    512 &  0.183 & 15.82 &  1.40 &  0.006 &  1.92 &  & 3668 &  0.176 & 20.41 &  1.45 &  0.061 &  1.89 \\
    788 &  0.303 & 15.50 &  0.84 &  0.261 &  5.58 &  & 3751 &  0.535 & 18.05 &  0.85 &  0.036 &  1.14 \\
    830 &  0.131 & 31.96 &  1.26 &  0.421 &  9.15 &  & 3810 &  0.076 & 10.71 &  0.87 &  0.006 &  1.35 \\
    906 &  0.398 & 15.73 &  1.04 &  0.056 &  2.70 &  & 3867 &  0.939 & 22.81 &  0.88 &  0.021 &  1.02 \\
   1102 &  0.187 & 18.34 &  1.11 &  0.041 &  1.68 &  & 3940 &  0.576 & 17.16 &  1.06 &  0.006 &  1.02 \\
   1112 &  0.442 & 32.14 &  1.39 &  0.246 &  2.16 &  & 5206 &  0.304 & 29.00 &  1.78 &  1.051 &  7.95 \\
   1154 &  0.193 & 18.59 &  1.69 &  0.006 &  0.69 &  & 5469 &  1.504 & 27.83 &  0.87 &  0.011 &  0.63 \\
   1223 &  0.974 & 10.08 &  0.67 &  0.021 &  0.87 &  & 5499 &  0.624 & 18.02 &  0.72 &  0.011 &  1.32 \\
   1308 &  0.215 & 10.71 &  0.86 &  0.021 &  1.83 &  & 5528 &  0.855 & 19.96 &  1.34 &  0.011 &  0.72 \\
   1463 &  0.192 & 21.99 &  0.81 &  0.051 &  2.40 &  & 5529 &  1.015 & 25.29 &  1.21 &  0.016 &  0.90 \\
   1694 &  0.388 & 17.83 &  1.68 &  0.006 &  0.72 &  & 5547 &  0.896 & 23.55 &  1.17 &  0.006 &  0.48 \\
   2126 &  0.401 & 25.30 &  0.99 &  0.006 &  0.75 &  & 5592 &  0.475 & 10.13 &  0.96 &  0.006 &  0.66 \\
   2132 &  0.090 & 10.70 &  0.60 &  0.096 &  6.03 &  & 5633 &  0.253 & 15.98 &  1.55 &  0.006 &  0.84 \\
   2201 &  0.380 & 11.10 &  0.76 &  0.121 &  3.21 &  & 6123 &  0.186 & 48.11 &  1.39 &  0.301 &  7.50 \\
   2206 &  0.602 & 11.80 &  0.85 &  0.016 &  0.81 &  & 6145 &  0.460 & 18.65 &  1.18 &  1.076 & 10.95 \\
   2220 &  0.951 & 14.06 &  0.79 &  0.006 &  0.90 &  & 6215 &  0.640 & 13.28 &  1.09 &  0.011 &  0.75 \\
   2268 &  1.954 & 21.63 &  0.82 &  1.496 &  8.19 &  & 6230 &  1.280 & 12.04 &  0.71 &  0.201 &  5.46 \\
   2377 &  0.496 & 41.49 &  1.25 &  0.091 &  2.64 &  & 6265 &  0.199 & 27.99 &  0.83 &  0.271 &  7.26 \\
   2502 &  0.512 & 13.24 &  0.91 &  0.071 &  1.47 &  & 6299 &  0.202 & 18.20 &  1.17 &  0.041 &  1.95 \\
   2512 &  0.359 & 19.42 &  0.96 &  0.176 &  4.23 &  & 6347 &  0.448 & 11.25 &  0.88 &  0.006 &  0.69 \\
   2564 &  0.256 & 10.33 &  0.83 &  0.011 &  1.02 &  & 6372 &  0.768 & 14.55 &  0.81 &  0.051 &  1.35 \\
   2597 &  0.256 & 10.90 &  1.02 &  0.006 &  0.78 &  & 6385 &  0.896 & 14.02 &  0.74 &  0.026 &  3.00 \\
   2632 &  1.443 & 22.19 &  0.78 &  0.006 &  0.63 &  & 6386 &  0.832 & 17.96 &  0.96 &  0.006 &  0.69 \\
   2649 &  0.256 & 14.79 &  1.23 &  0.011 &  0.90 &  & 6569 &  0.936 & 10.45 &  0.78 &  0.006 &  0.81 \\
   2755 &  0.144 & 11.10 &  0.74 &  0.171 &  5.04 &  & 6573 &  0.289 & 12.75 &  0.99 &  0.011 &  0.99 \\
   2788 &  0.872 & 25.63 &  0.87 &  0.011 &  0.84 &  & 6591 &  0.773 & 22.82 &  1.01 &  0.011 &  0.81 \\
   2861 &  1.599 & 12.87 &  0.73 &  0.071 &  1.92 &  & 6689 &  0.192 & 16.38 &  0.88 &  0.021 &  1.41 \\
   2892 &  0.288 & 11.95 &  0.90 &  0.051 &  1.77 &  & 6693 &  0.283 & 20.10 &  1.37 &  0.036 &  1.38 \\
   2896 &  0.456 & 51.35 &  0.72 &  0.276 &  5.34 &  & 6700 &  0.192 & 15.33 &  1.42 &  0.461 &  8.58 \\
   2918 &  0.448 & 16.47 &  0.87 &  0.056 &  2.13 &  & 7060 &  0.146 & 14.04 &  0.74 &  0.071 &  3.06 \\
   2952 &  0.680 & 35.08 &  1.29 &  0.021 &  1.47 &  & 7173 &  0.966 & 22.50 &  1.20 &  0.006 &  0.57 \\
   2977 &  0.576 & 15.12 &  1.18 &  0.016 &  0.93 &  & 7227 &  0.092 & 12.30 &  1.07 &  0.156 &  3.84 \\
   3066 &  0.176 & 16.18 &  1.15 &  0.011 &  0.93 &  & 7292 &  0.262 & 26.31 &  1.06 &  0.396 &  6.24 \\
   3078 &  0.224 & 17.94 &  1.47 &  0.006 &  0.84 &  & 7813 &  0.564 & 32.57 &  0.97 &  0.371 &  3.57 \\
   3121 &  0.792 & 11.76 &  0.90 &  0.006 &  0.60 &  & 7939 &  1.039 & 57.77 &  0.79 &  0.006 &  0.27 \\
   3282 &  0.078 & 14.13 &  1.15 &  0.021 &  2.07 &  & 7970 &  0.387 & 11.37 &  0.89 &  0.006 &  0.87 \\
   3323 &  0.448 & 14.77 &  0.92 &  0.006 &  0.72 &  & 7988 &  0.413 & 16.21 &  0.83 &  0.131 &  4.62 \\
   3338 &  0.108 & 12.32 &  1.03 &  0.111 &  6.33 &  & 7995 &  0.608 & 11.52 &  0.68 &  0.026 &  1.38 \\
   3340 &  1.016 & 16.30 &  1.04 &  0.016 &  0.99 &  & 8047 &  0.892 & 32.19 &  1.33 &  0.021 &  0.60 \\
   3359 &  0.344 & 25.65 &  1.19 &  0.056 &  1.59 &  & 8076 &  0.218 & 12.31 &  0.82 &  0.226 &  6.57 \\
   3379 &  0.608 & 22.48 &  0.98 &  0.006 &  0.60 \\
\hline
\end{tabular}
\caption{Sample of the short GRBs observed by BATSE. The bursts were selected on the
basis of their signal to noise ratio, $S/N>10$, in the energy band 50-300 keV.
The total sample contained 182 bursts; the table lists only these bursts
that fitted to the defined profile shape (Eq. \ref{eq:funsvensson}). 
\label{tab:batse}}
\end{table*}

We fit the pulse profiles to the shape defined by the equation
\ref{eq:funsvensson}.  The data are in units of [cts/s] and
we do not convert them to the photon flux, due to the
large error bars in case of short bursts. The conversion to physical 
units would dramatically increase the uncertainty.

\begin{figure}
\epsfxsize = 250pt
\epsfbox{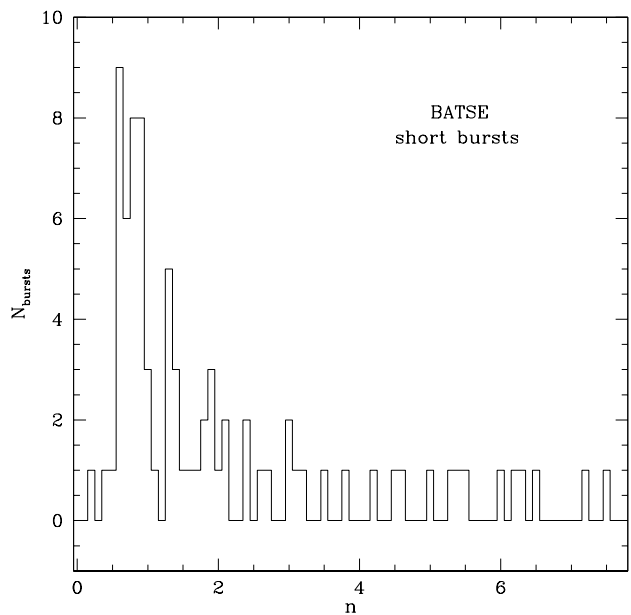}
\caption{The distribution of the index $n$, fitted to the profiles
of the short GRBs observed by BATSE.
\label{fig:wyk2batse}}
\end{figure}

In the Figure \ref{fig:wyk2batse} we show the distribution of the 
index $n$ in the observed sample.
The plot contains only these bursts, that fitted to the defined profile shape.
Only less than half of the originally selected bursts remained (83 out of 182).

As the Figure shows, no significant excess in the distribution
is seen around the index $n=3$. However, an excess at low values of $n$ is
seen quite clearly. The figure shows overall similarity to the theoretical
prediction based on thick shell approximation while it is definitively different
from Figure~\ref{fig:wyk2}.

\subsection{Statistical distribution of bursts}
\label{sec:statistics}

Below we show the results of the Monte Carlo calculations of the statistical 
distribution of GRB durations. In all the simulations we take $N_{\rm obs}=1000$
observers (or, equivalently, 1000 GRBs at different positions on the sky with 
respect to the observer). Each of the bursts consists of
$N_{\rm sub}=350$ sub-jets,  which have the opening angles of 
$\Delta \theta_{\rm sub} = 0.02$ and Lorentz factor $\gamma=100$.
 The departure time of each sub-jet, $t_{\rm dep}$,
is chosen randomly within the total activity time of the central engine, i.e. 
between 0 and $T_{\rm dur}=30$ sec. The sub-jets have a quasi-Gaussian distribution 
within the jet of the total opening angle $\Delta \theta_{\rm tot} = 0.2$.

We compare the results for two cases:
the instantaneous emission form a thin shell at $r=R_{0}$ and
the emission from the thick expanding shell, from $r=R_{\rm min}$ to 
$r=R_{\rm max}$. 
The infinitesimally thin emitting shell is located at  $r_{0} = 10^{14}$ cm,
and the thick shell emits the radiation from $R_{\rm min} = 10^{14}$ cm, to 
$R_{\rm max} = 2\times 10^{14}$ cm.
Below, in the Section, \ref{sec:onaxis},
 we  discuss the distribution limited to 
these bursts that contain only the sub-jets, and in Section \ref{sec:offaxis}
we show the results for the bursts that include the emission from
every sub-pulse, seen either on axis or off-axis. The criterion for
counting these sub-bursts as a part of the GRB is their flux
above a background for the duration time $T_{90}$,
as observed by the BATSE detector.

\subsubsection{Statistics for nearly-on-axis observers}
\label{sec:onaxis}

As a first step, we will study the distribution of burst durations
under the (restrictive) assumption that only the sub-jets that are on
 the line of 
sight contribute to the GRB event. In other words, we neglect any emission 
that is observed off-axis, which could possibly contribute to gamma rays, 
assuming that those events would only give rise to the X-ray Flashes.
We require that all the sub-jets that are contributing to the GRBs
will satisfy the condition \ref{eq:inside} with respect to the observer axis.

\begin{figure}
\epsfxsize = 250pt
\epsfbox{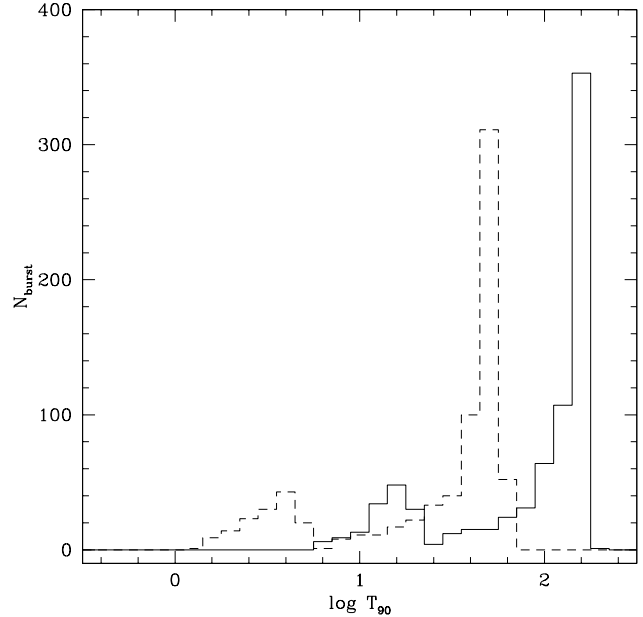}
\caption{The simulation 
with $N_{\rm obs} =1000$ random observers, and the jet
consisting of $N_{\rm sub}=350$ sub-jets. The histogram 
shows the number of bursts (observer's
positions) in the function of the total burst duration time.
Each burst includes only the sub-jets caught on the line of sight.
The emission is produced either by an extended shell from $R_{\rm min} = 10^{14}$ cm, to 
$R_{\rm max} = 2\times 10^{14}$ cm (solid line) or by
the thin shell at $r_{0} = 10^{14}$ cm (dashed line).
\label{fig:Hist1}}
\end{figure} 

The  observer may catch a sub-jet on his line of sight either
while looking very close to the main jet axis, with $\theta_{\rm obs}\sim 0$ 
(multiple sub-jets are seen), 
or while looking close to the edge.  In the first
 case, the number of detected sub-jets 
is about 300, and
in the latter case, the observer 
is most likely to catch one sub-jet (about 10\% of observers). 
Obviously, the largest number of observers do not catch any
single sub-jet.

In the Figure \ref{fig:Hist1} we show the histogram of burst durations, 
for this simulation. The distribution has two peaks, 
located around 4 and 50 seconds, or 8 and 65 seconds, for an
instantaneous or extended emission from the sub-jets, respectively. 
The exact location of the peaks depend on the model parameters,
and in particular the short GRB peak is sensitive to the value of the radius
of emitting shell, as well as the opening angle of a sub-jet (cf. Eq. \ref{eq:tmin}).
For example, $\Delta\theta_{\rm sub} = 0.015$ shifts the short burst duration peak
to $T_{90} = 2.5$ s, while the radius of $r_{0} = 10^{13}$ cm shifts
this  peak to 
$T_{90} = 0.4$ s.
The long burst peak does not depend on these parameters, but rather 
 on the duration time of the central engine activity.
If we take $T_{\rm dur}=0.3$ s, instead of 30 s, the peak for long bursts is 
shifted to $T_{90}=8$ s. However, the peak for short GRBs does not change
and is located 
at $\sim 4$ s, so
the shorter activity time results in the essentially single-peak distribution
for only short bursts.


\subsubsection{Statistics for the off-axis emission included}
\label{sec:offaxis}

In this section we discuss the effect of the off-axis emission that
contributes to gamma rays. Since the observed flux drops very rapidly 
with the increasing inclination of the line of sight, as soon as the observer axis is located outside the sub-jet cone, it is crucial to
determine the condition for the GRB to be observed. This is  connected 
mainly with the flux limit of the detector.

For example, BATSE will generate a burst trigger if the count rate in two or more
detectors exceeds a threshold specified in units of standard deviations
above background (nominally 5.5). The actual detection depends only slightly on the 
spectral hardness

\begin{figure}
\epsfxsize = 250pt
\epsfbox{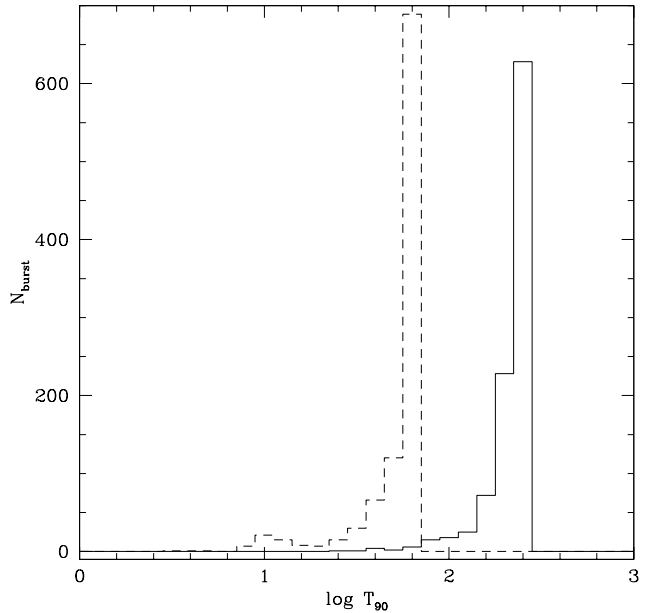}
\caption{The results of the simulation with the emission from off-axis sub-jets
included.  The bursts include only
these off-axis sub-bursts, for which 
$F^{peak}_{\rm 10 keV} > 5.5\times \sqrt{F_{\rm B}/T_{90}}$. 
Other parameters are the same as in Fig. \ref{fig:Hist1}. 
\label{fig:Hist2}}
\end{figure}

In the Figure \ref{fig:Hist2} we show the results of the simulation.
The off-axis emission of the sub-jets contributes
 to GRB whenever 
\begin{equation}
F^{peak}_{10 keV}>5.5\times \sqrt{F_{\rm B}/T_{90}}.
\label{eq:fluxlimit}
\end{equation} 
where the background flux is arbitrarily chosen to be $F_{\rm B}=1.0$

The shortest GRBs are mainly due to the 
bursts that include only the off-axis emission.
 This is because the weak
off-axis sub-bursts in general tend to have longer durations, 
 which sometimes prevents them from falling below the threshold.
On the other hand, the weakest of these sub-bursts have too small
flux at 100 keV, and are missed anyway.
Therefore the total number of the sub-bursts that contribute
to the left wing of this histogram is small and
the only sub-bursts that are counted here are the off-axis sub-bursts.

The sub-bursts caught on-axis (or nearly on-axis)
give the main contribution to the longest GRBs. Since their
emitted flux is always large, the threshold limit does not affect them.
These sub-bursts are seen mostly at very small observing angle $\theta_{\rm obs}$,
and a large number of them is detected. This means a large spread between the 
starting time of the first sub-jet and ending time of the last one,
which translates into a long total event.

The shape of the histogram is clearly not bimodal.
A small excess of short bursts around $T\sim 10$ s appears only for the model with
instantaneous emission at $r=r_{0}$, however this
feature disappears already 
in the model with expanding shell emission from $r_{\rm min}$ to 
$r_{\rm max}$.
The bimodal distribution is found by 
Toma et al. (2005) in the model with only
 instantaneous emission at $r=r_{0}$.
However, these authors consider the case when the events are classified by 
the detector as a GRB on the basis of their spectral 
hardness. They select only these events
for which
the hardness ratio, i.e. the ratio of fluency defined as $
S_{\nu_{1}, \nu_{2}} = \int_{T_{start}}^{T_{\rm end}} \int_{\nu_{1}}^{\nu_{2}} 
F_{\nu} d\nu dT$, is smaller than  
$S_{\rm 2 keV, 30 keV}/S_{\rm 30 keV, 400 keV} \approx 0.3$.
 In this case  the contribution from the
short bursts is much more pronounced.

\subsubsection{Distribution of redshift}

The distribution of the long bursts with redshift is now thought to 
trace the  star 
formation in an unbiased way (Jakobsson et al. 2005).
Therefore the probability of finding a bursts with a redshift z is
given by:
\begin{equation}
dP(z) = A D^{2} R(z) dz {dr \over dz}
\end{equation}
where $R(z)$ is the star formation rate in the comoving frame (Porciani \& Madau 2001), $D$ is the cosmological distance (Chdorowski 2005), $A$ is the normalization constant and $dr/dz = c/H_{0}/\sqrt{\Omega_{\rm M}(1+z)^{3}+\Omega_{\Lambda}}$. We adopt the cosmological parameters of $\Omega_{\rm M}=0.3$, $\Omega_{\Lambda}=0.7$ and $H_{0}=65$. 
The normalization is calculated from the condition that:
\begin{equation}
\int_{z_{\rm min}}^{z_{\rm max}} P(z) dz = 1.
\end{equation}

\begin{figure}
\epsfxsize = 250pt
\epsfbox{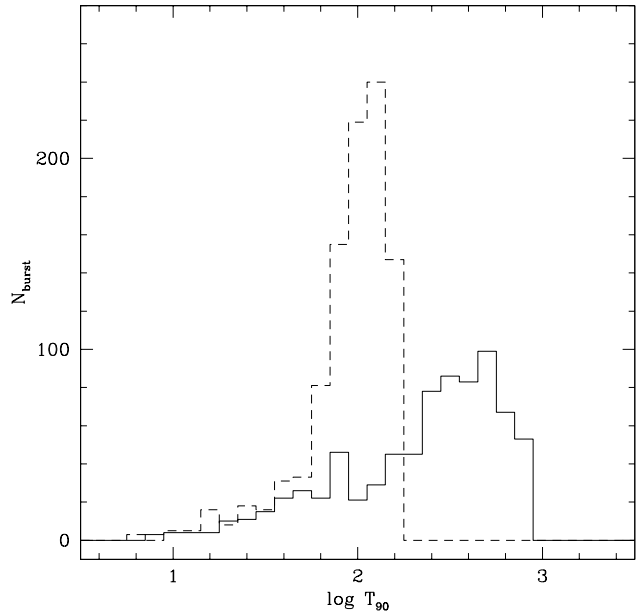}
\caption{The same as in Fig. \ref{fig:Hist2}, but with a random redshift 
distribution included. The redshift for each burst was
 chosen between 0 and 4, with a probability distribution that traces the star 
formation rate, $R_{\rm SF}(\Omega_{\rm M}=0.3, \Omega_{\rm \Lambda}=0.7)$. 
\label{fig:Hist3}}
\end{figure}

Here we check the influence of the redshift distribution on our model.
For each of the 1000 bursts we randomly choose its redshift, within a
distribution with a cut-off at $z_{\rm max} = 4$. The Figure \ref{fig:Hist3} shows 
the resulting statistics of bursts.

The duration distribution is now more smeared than in case of a constant redshift, 
$z=1.0$, and shifted toward longer durations. 
The total number of bursts that satisfy the condition (\ref{eq:fluxlimit})
is similar to that in Fig. \ref{fig:Hist2} and about $\sim 700$ out of a 1000 
bursts exhibit a large  flux in gamma rays. 
However, the durations of the bursts
with the distribution of redshifts tend to be more equal, due to the 
linear dependence on $1+z$. This makes the peak of the histogram flatter.
The shift toward longer durations results from the fact 
the redshift distribution traces the star formation rate according to the function 
$Rsf_{2}$ in Porciani \& Madau (2001) so the number of events
with $z<1.0$ is very small while saturates for large $z$.
Therefore the
long bursts can be longer, due to the contribution of 
bursts with $z>1.0$, while the short bursts have similar durations to these
in Fig. \ref{fig:Hist2}.

\section{Discussion and conclusions}
\label{sec:diss}

The scenario in which the central engine of a GRB operates like a ``shot-gun''
emitting multiple narrow bullets was
originally suggested by Heinz \& Begelman (1999). 
Also, Fenimore et al. (1999) showed that the variability of GRB may be due to
the breaking of the local spherical symmetry of the emitting shells.
In this case the duration of 
the total event was determined by the time of activity of the central engine,
and therefore connected with the progenitor type.
The observational evidence for the association of the long
GRBs with supernovae (Bloom et al. 2002; Stanek et al. 2003)
support the collapsar model. The simulations of the
jet propagating through the envelope of the collapsing star 
(e.g. Aloy et al. 2000; MacFadyen et al. 2001)  
 show
that the jet has an ultimate Lorentz factor of about 100-150 
and an opening angle wider than $\sim 10^{\circ}$. The jet does not have a
uniform structure in its density, velocity, pressure etc. These physical 
quantities have a certain distribution along with the angular distance from the jet 
axis, however, the gradients are not very sharp, and
it is not certain if this scenario could result in a separate ``bullets''
released from the center during the time of activity. 

One of the possible scenarios is the quasi-universal structured jet 
model (e.g. Rossi et al. 2002; Zhang \& Meszaros 2002; Lloyd-Ronning et
al. 2004), in which the observed diversity of GRB properties is
explained by a different viewing angle. The jet has a standard geometrical 
configuration, with a stable outflow channel but jet properties 
(e.g. Lorentz factor, energy flux) are 
distributed non-uniformly per solid angle within the cone.

This is in contrast with the
simple jet models, in which the energy distribution within the cone
was uniform (Rhoads, 1997; Panaitescu et al. 1998).
These simple jets have been used to successfully explain the breaks in the GRB
optical afterglow lightcurves (Moderski et al. 2000; Huang et al. 2000; 
Granot et al. 2002).
Since the energy is collimated to different degrees in various bursts,
the jet opening angle is the main
parameter that drives the jet properties. Which mechanism drives the various 
collimation degree in various GRBs remains to be explained.

The model studied in the present work
invokes a kind of a ``shot-gun'' model, with a fixed opening angle,
and a Gaussian distribution of the subjet probability around the symmetry
axis. In the case of assumed instantaneous emission the model is effectively 
quite close to a ``shot-gun'' scenario while in the case of extended emission
(i.e. thick shell approximation) the emitting gas has definitively a jet-like
appearance. However, the model is different from the single structured jet 
since the distribution of the  directions of sub-jets contains the element
of randomness. The underlying assumption is that the process of the jet 
formation is not stable, and an open jet channel may close up and be replaced
with a new one. In this model the GRB properties depend significantly
on the observer's position.

We first analyzed the profiles of single pulses coming from separate
sub-jets. 

Observationally, such profiles were studied by Ryde \& Svensson 
(2002) for $T_{90} > 2$ s bursts in BATSE sample. 
They found a good representation of
single subpulse profiles with equation (\ref{eq:funsvensson}), and noticed
a two-peak distribution of the index $n$: a higher peak around $n=1$ and a
lower peak around $n=3$. We completed their analysis for $T_{90} < 2$ and
found one peak below $n=1$ but no significant traces of the second one.
If we restricted ourselves to the data with the highest S/N ratio (above 30) 
perhaps we could
see a slight relative enhancement at $n \sim 3$, but we have only 11 such 
events in the sample.

We compared these results with the theoretical profiles. 
Assuming instantaneous emission we
obtained values of $n$ between $n=4$ and$n=6$ . However, the 
emission from thick shell with $R_{\rm min}=10^{14}$ cm and
$R_{\rm max}=2\times 10^{14}$ cm results in pulse profiles
that are slightly wider in their rising part and have a different decay 
profiles. The corresponding distribution of the index $n$ is much broader
and peaks below $n=1$. Therefore, the thick shell approximation seems to be
a better representation of the observational data within the frame of the 
adopted model.

We next analyzed the question whether the adopted model can explain the
apparent existence of the two classes of bursts: short and long.

 We found that the multiple sub-jet model can reproduce the bimodal 
distribution of GRB
durations only for some specific conditions and parameters.
In particular, this kind of distribution is reproduced if
only  the sub-jets  that are seen on the line of sight 
can contribute to the GRB. When this assumption is released and all the 
sub-jets
can contribute to the GRB (provided they give a sufficient flux in gamma-rays),
the bimodal distribution is
smeared.
 Furthermore, the bimodal distribution vanishes completely whenever
we assume the emission from a thick shell from
$R_{\rm min}$ to $R_{\rm max}$ instead of instantaneous emission at $R_{0}$

In the sub-jet model studied in this paper
the duration time of the GRB depends on two 
parameters: (i) the radius of the emitting photosphere of a 
shell, $r_{0}$ (or, more specifically, its combination with the
 sub-jet opening angle, $\Delta\theta_{\rm sub}$)
and (ii) on the time of activity of the central engine, $T_{\rm dur}$.
 The case (i) refers to the single sub-pulses detected by the observer, 
while the case (ii) is for the multiple sub pulses.

An important ingredient of the model is the multiplicity 
of the sub-jets. This is parameterized by the ratio
of the solid angle  covered by all the 
sub-jets to the solid angle covered by the whole jet:
$\xi=(N_{\rm sub} \times \Delta \theta_{\rm sub}^{2})$/$\Delta \theta_{\rm tot}^2$.
If this ratio is larger than unity, i.e. we have multiple sub-jets
scattered with some, say, Gaussian, distribution around the jet axis,
most of the bursts contain multiple pulses and the chance of detecting 
a single pulse is large only very close to the edge of the jet.
For the standard parameters used throughout this article we
had $\xi = 350\times(0.02)^{2}/(0.2)^{2}=3.5$ 
The duration of the GRB is governed then by the activity time $T_{\rm dur}$,
and the simulations show that the bimodal distribution is 
possible only for two distinct values
of this activity time.

\begin{figure}
\epsfxsize = 250pt
\epsfbox{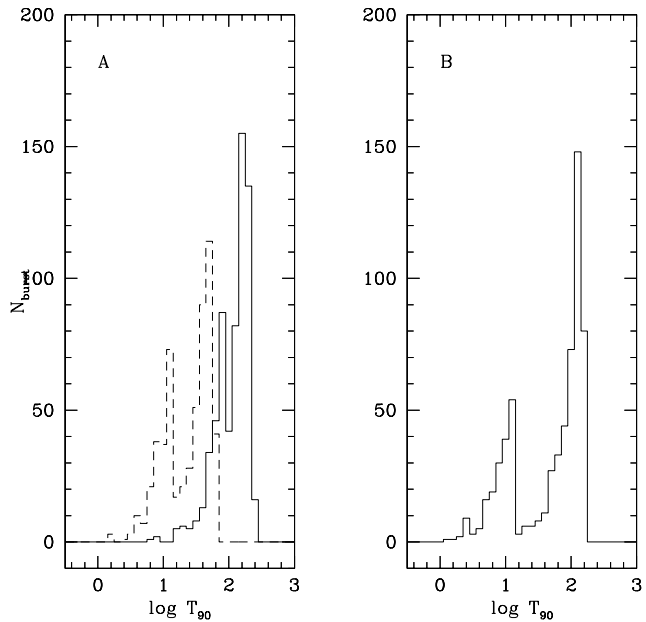}
\caption{Histograms for models with very small sub-jet multiplicity: $N_{\rm sub}=10$,
$\Delta \theta_{\rm sub}=0.02$, $\Delta \theta_{\rm tot}=0.2$. A: emission from 
$r_{0}=10^{14}$ cm (dashed line) and from $R_{\rm min}=10^{14}$ cm to  
$R_{\rm max}=2 \times 10^{14}$ cm (solid line); B: emission 
from $R_{\rm min}=1.66 \times 10^{13}$ cm to  $R_{\rm max}=3.33 \times 10^{13}$ cm.
\label{fig:nsub10}}
\end{figure}

We have a different situation  for $\xi \ll 1$. Most of the bursts will then
contain only single pulses, and their duration is governed by the
photospheric radius $r_{0}$. We checked, that for extremely low multiplicity
($N_{\rm sub}=10$, which gives $\xi = 0.1$) the bimodal distribution duration can be
recovered, both for instantaneous and extended shell emission (see Figure 
\ref{fig:nsub10}).
However, a question arises, if in such a case we can really
talk about the ``multiple sub-jet model''. Also, the observed GRB pulse
profiles would look totally different, as we would have mostly single or double spikes
even for observers looking very close to the jet axis. The observations of long GRBs
show something
different: many of them exhibit a narrow substructure 
rather than couple of spikes.

To sum up,  we suggest that an explanation of the bimodal duration 
distribution of the observed GRBs
is possible rather on the basis of two independent kinds of sources responsible
for short and long events.
This is further justified by the fact that
 apart from many similarities, there are also systematic
differences between long and short bursts, such as their spectral hardness or
the systematically lower fluence in the case of short bursts.

\section*{Acknowledgments}
We thank Marek Sikora, Tomek Bulik and Micha{\l} Chodorowski 
for helpful discussions.
This work was supported in part by grant No. PBZ 057/P03/2001
 of the Polish Committee for Scientific Research.

\end{document}